\documentclass[epsfig,graphicx,16pt]{article} 
\usepackage{graphicx}
\usepackage{amsmath}
\usepackage{mathtools}
\usepackage{blindtext}
\newcommand{\ackname}{Acknowledgements}

\begin{document}

\title{
{\it Two-Point Evolution Equations for Incompressible Variable Density Turbulence.
}}
\author{Timothy T. Clark \\The University of New Mexico \\Department of Mechanical Engineering \\Albuquerque, New Mexico}
\date{\today}

\maketitle

\begin{abstract}
A derivation of the ``exact" two-point equations analogous to those used as a basis for one-point Reynolds-Averaged Navier-Stokes turbulence model for variable density, incompressible turbulence. The purpose is to present the statistical correlations that must be dealt with to yield a two-point, or alternatively, a spectral model of variable density turbulence.
\end{abstract}


\section{Two-Point Correlation Equations}\label{Sec:TwoPoint}
The two-point velocity covariance tensor is
\begin{equation} \label{Eqn:TwoPoint_1}
T_{ij}\left({\bf x_1},{\bf x_2}\right) = \overline{u'_i \left({\bf x_1}\right) u'_j\left({\bf x_2}\right)},
\end{equation}
where the overline denotes and ensemble average and the prime denotes a fluctuation about the volume-weighed  ensemble average. We have dropped the time argument solely for brevity. This definition of the two-point covariance is symmetric under the simultaneous interchange of coordinate and index.
Using the variable substitution suggested by Besnard {\it et al.}, \cite{BHRZ1},
\begin{equation}
\begin{aligned}
{\bf x} &= \frac{1}{2}\left({\bf x_1}+{\bf x_2}\right), \\
{\bf r} &= \left({\bf x_1}-{\bf x_2}\right),
\end{aligned}
\end{equation}
the two-point correlation becomes,
\begin{equation}
T_{ij}\left({\bf x},{\bf r}\right) = \overline{u'_i \left({\bf x}+\frac{1}{2}{\bf r}\right) u'_j\left({\bf x}-\frac{1}{2}{\bf r}\right)}.
\end{equation}
If the turbulence is homogeneous, then the two-point velocity covariance is invariant under translation in ${\bf x}$ and is simply a function of the separation vector, ${\bf r}$;
\begin{equation}
T_{ij}\left({\bf r}\right) = \overline{u'_i \left({\bf x}+\frac{1}{2}{\bf r}\right) u'_j\left({\bf x}-\frac{1}{2}{\bf r}\right)},
\end{equation}
and if the turbulence is isotropic, it is invariant under rotations and the covariance is simply a function of the separation distance, $r = \left|{\bf r}\right|$
\begin{equation}
T_{ij}\left(r\right) = \overline{u'_i \left({\bf x}+\frac{1}{2}{\bf r}\right) u'_j\left({\bf x}-\frac{1}{2}{\bf r}\right)}.
\end{equation}
Batchelor \cite{BatchelorBook1} has elucidated many of the properties of tensors under the conditions of statistical homogeneity.  The arise, in part, due to the symmetry properties of equation \ref{Eqn:TwoPoint_1}.  

In developing a set of two-point equations, Clark and Spitz \cite{CSModel} sought to construct an appropriate definition of the two-point velocity covariance tensor that satisfied the symmetry constraints of equation \ref{Eqn:TwoPoint_1}, but was consistent with the definition of the mass-weighted Reynolds-stress tenor used in Los Alamos National Laboratory's BHR models \cite{BHRZ2,BHR3TLSM}.  Clark and Spitz chose
\begin{equation} \label{Eqn:TwoPoint_2}
R_{ij}\left({\bf x_1},{\bf x_2}\right) = \frac{1}{2}\overline{\left(\rho \left({\bf x_1}\right)+\rho \left({\bf x_2}\right)\right)u''_i \left({\bf x_1}\right) u''_j\left({\bf x_2}\right)},
\end{equation}
where $\rho$ is the fluid density and the double-primes denote a fluctuation about the mass-weighted ensemble average of the velocities.  We will not dwell on the intricacies of mass-weighted averages, but the interested reader is directed to Besnard {\it et al.} \cite{BHRZ2} and Schwarzkopf {\it et al.} \cite{BHR3TLSM}. Note that in the one-point limit (${\bf x_1} \rightarrow {\bf x_2} \rightarrow {\bf}$), we recover the single point mass-weighted Reynolds stress tensor.
Clark and Spitz used the same variable substitutions as Besnard  {\it et al.}, and went on to construct the two-point equations corresponding to the one-point equations developed by Besnard {\it et al.} (1992), except for the energy equation.  This current effort has reexamined the exact (unmodeled) correlation equations constructed by Clark and Spitz (1995), and corrected an error in that earlier work.  The present equations, documented in the Appendix, govern the evolution of the two-point Reynolds stress tensor, equation \ref{Eqn:TwoPoint_2}.  The dominant mean-pressure coupled production term in the two-point Reynolds-stress tenor is the two-point turbulent mass flux, $a_i\left({\bf x_1},{\bf x_2}\right)$,
\begin{equation} \label{Eqn:TwoPoint_a}
\begin{aligned}
a_{i}\left({\bf x_1},{\bf x_2}\right) &= -\overline{\frac{\left[\rho\left({\bf x_1}\right) + \rho\left({\bf x_2}\right)\right]}{2}\left\{
u''_i\left({\bf x_1}\right)\left[v\left({\bf x_2}\right) -\frac{1}{\overline{\rho}\left({\bf x_2}\right)}\right]\right\}}\\
&= -\frac{1}{2}\left\{
\overline{\rho\left({\bf x_1}\right)u''_i\left({\bf x_1}\right)v\left({\bf x_2}\right)}
+\overline{u''_i\left({\bf x_1}\right)}
-\frac{\overline{ \rho\left({\bf x_2}\right)u''_i\left({\bf x_1}\right)}}{\overline{\rho}\left({\bf x_2}\right)}
\right\},
\end{aligned}
\end{equation}
This term differs from that found by Clark and Spitz, although both this correlation and that of Clark and Spitz both collapse in the one-point limit to the turbulent mass-flux correlations used by Besnard {\it et al.} (1992) and Schwarzkopf {\it et al.} (2014).
The dominant mean-pressure coupled production term in the evolution equation for two-point mass-flux, $a_i\left({\bf x_1},{\bf x_2}\right)$, is the two-point density-specific volume correlation, $\beta\left({\bf x_1},{\bf x_2}\right)$, which is related to the $b$-parameter in the BHR \cite{BHRZ2} and BHR-3 \cite{BHR3TLSM} models
\begin{eqnarray}\label{Eqn:TwoPoint_b}
\beta\left({\bf x_1},{\bf x_2}\right)
= \frac{1}{2}\overline{\left[\rho\left({\bf x_1}\right) + \rho\left({\bf x_2}\right)\right]
\left[v\left({\bf x_1}\right) - \frac{1}{\overline{\rho}\left({\bf x_1}\right)}\right]\left[v\left({\bf x_2}\right) - \frac{1}{\overline{\rho}\left({\bf x_2}\right)}\right] },
\end{eqnarray}
where $v = 1/\rho$ is the specific volume.  In the one-point limit, this is related to the $b$-parameter by
\begin{eqnarray}\label{Eqn:OnePoint_b}
\beta\left({\bf x},{\bf x}\right)
= - \frac{ \overline{ \rho'\left({\bf x}\right)v'\left({\bf x}\right) } }{\overline{\rho}\left({\bf x}\right)} = \frac{b\left({\bf x}\right)}{\overline{\rho}\left({\bf x}\right)}.
\end{eqnarray}
The development of the exact correlation equations  along with a brief discussion is presented in the Appendix.  It is hoped that they will provide guidance in the analysis of Direct Numerical Simulations of variable density turbulence, as well as in the development of more advanced models.

\section*{\ackname}
This work was performed under LANL Subcontract 325696.  Dr. Clark thanks Drs. Michael Steinkamp and Robert Gore of Los Alamos National Laboratory.


\clearpage
\appendix{Appendix: The Two-Point ``BHR" equations for Variable Density Turbulence}
\section{Introduction and Purpose}
The derivation of the Reynolds-averaged Navier-Stokes turbulence equations for variable density turbulence is a tedious task, requiring meticulous derivation and copying.  The resulting equations may be recast in a variety of ways to take advantage of certain correlations, or to exploit conservative forms of specific terms.  Additionally, the use of mass-weighted averages and fluctuations requires choosing between modeling particular expressions in terms of the mass-weighted statistics or the volume-weighted terms.  We point this out simply so that the reader is aware that the following derivations are not the only way to derive or represent these two-point equations.  However, the purpose of this derivations is two-fold: (1) to demonstrate the BHR-like hierarchy as represented in the two-point equations (and thus to correct an apparent error in a previous work), and (2) to provide guidance to those conducting experiments and direct numerical simulations regarding what may be preferred two-point statistics to measure or calculate for variable density turbulence, rather than using the metrics derived for constant-density incompressible turbulence.  

The following collection of derivations is ``archival"--far more detail is included than would be appropriate for a peer-reviewed journal publication.  However, the included detail permits the interested researcher to easily verify (or, possibly correct) the derivations, and to leave no questions as to how the equations were derived.

\section{Governing Equations}
The derivation begins with the  Navier-Stokes equations;
\begin{eqnarray}\label{Eqn:NS}
\frac{\partial \rho u_i}{\partial t} +\frac{\partial \rho u_i u_n}{\partial x_n} =  \frac{\partial \sigma_{in}}{\partial x_n}
\end{eqnarray}
where
\begin{eqnarray}
\sigma_{ij} =\tau_{ij} - \delta_{ij} p,
\end{eqnarray}
and
\begin{eqnarray}
\tau_{ij} = \mu \left(\frac{\partial u_i}{\partial x_j} +\frac{\partial u_j}{\partial x_i} -\frac{2}{3}\delta_{ij} \frac{\partial u_n}{\partial x_n}\right) .
\end{eqnarray}
The continuity equation is
\begin{eqnarray}\label{Eqn:Continuity}
\frac{\partial \rho}{\partial t} +\frac{\partial \rho u_n}{\partial x_n} = 0,
\end{eqnarray}
the species mass-fraction equations are
\begin{eqnarray}\label{Eqn:Concentration}
\frac{\partial \rho c_k}{\partial t} +\frac{\partial \rho u_nc_k}{\partial x_n} = \frac{\partial}{\partial x_n}\left\{\rho {\cal D}\frac{\partial c_k}{\partial x_n}\right\},
\end{eqnarray}
and the energy equation is
\begin{eqnarray}\label{Eqn:Energy}
\frac{\partial \rho I}{\partial t} +\frac{\partial \rho u_n I}{\partial x_n} = \sigma_{nm} \frac{\partial u_n}{\partial x_m} + \frac{\partial }{\partial x_n}\left\{\kappa \frac{\partial T}{\partial x_n}\right\}.
\end{eqnarray}

\section{Averaged Equations}
The usual Reynolds decomposition represents a variable, e.g., the velocity vector $u_i$ as an ensemble-averaged part, denoted by a simple overbar, and a fluctuation, denoted by a prime. For example,
\begin{equation}
u_i = \overline{u}_i +u'_i, 
\end{equation}
where
\begin{equation}
\overline{u}_i = \frac{1}{N_e}\sum_{k=1}^{N_e} u^{\left(k\right)}_i,
\end{equation}
where $N_e$ is the number of realizations in the ensemble and the superscript $\left(k\right)$ denotes the $k^{th}$ element in the ensemble.  Thus the fluctuation is simply
\begin{equation}
u'_i = u_i -\overline{u}_i, 
\end{equation}
and
\begin{equation}
\overline{u'}_i =\left\{\frac{1}{N_e} \sum_{k=1}^{N_e} u^{\left(k\right)}_i \right\} -\overline{u}_i = 0.
\end{equation}
The mass-weighted (or ``Favre") average is denoted by a tilde, and again refers to an ensemble average;
\begin{equation}
\tilde{u}_i = \frac{1}{\overline{\rho} }\frac{1}{N_e}\sum_{k=1}^{N_e} \rho u^{\left(k\right)}_i = \frac{\overline{\rho u_i}}{\overline{\rho}},
\end{equation}
where again the overbar simply denotes an ensemble average.  Note that the mass-weighted fluctuation is denoted by a double-prime,
\begin{equation}
u''_i = u_i - \tilde{u}_i,
\end{equation}
and we note that multiply the above equation by $\rho$ and ensemble averaging shows that
\begin{equation}
\overline{\rho u''_i} = \overline{\rho u_i} - \overline{\rho}\tilde{u}_i =  \overline{\rho}\tilde{u}_i - \overline{\rho}\tilde{u}_i  =0.
\end{equation}
Note that the Favre-averaged velocity  is related to the Reynolds averaged velocity as follows;
\begin{equation}\label{Eqn:Favre-Reynolds1}
\overline{u}_i + u'_i = \tilde{u}_i + u''_i.
\end{equation}
Ensemble averaging the above equation gives
\begin{equation}
\overline{ u}_i + \overline{u'}_i= \tilde{u}_i + \overline{u''}_i,
\end{equation}
where we note that $\overline{u'}_i = 0$ so that
\begin{equation}\label{Eqn:Favre-Reynolds2}
\overline{ u}_i = \tilde{u}_i + \overline{u''}_i,
\end{equation}
Subtracting equation \ref{Eqn:Favre-Reynolds2} from \ref{Eqn:Favre-Reynolds1} yields,
\begin{equation}\label{Eqn:Favre-Reynolds3}
u'_i = u''_i +  \overline{u''}_i .
\end{equation}
Note that in the usual LANL``formalism," we define the turbulent mass-flux (a vector) as $a_i$;
\begin{equation}\label{Eqn:Favre-Reynolds4}
a_i =  \overline{u''}_i,
\end{equation}
so that equation \ref{Eqn:Favre-Reynolds3} becomes
\begin{equation}\label{Eqn:Favre-Reynolds5}
u'_i = u''_i - a_i .
\end{equation}
Finally note that multiplying equation \ref{Eqn:Favre-Reynolds5} through by the density and performing ensemble averages, we find that
\begin{equation}\label{Eqn:Favre-Reynolds6}
\overline{\rho u'_i}  = \overline{\rho  u''_i} - \overline{\rho a_i }  =  - \overline{\rho a_i },
\end{equation}
and 
\begin{equation}\label{Eqn:Favre-Reynolds7}
\begin{aligned}
\overline{\rho u'_i}  &= \overline{\left[\overline{\rho} + \rho'\right]u'_i} = \overline{\rho}\overline{u'_i} + \overline{\rho' u'_i}\\
&= \overline{\rho' u'_i},
\end{aligned}
\end{equation}
so that
\begin{equation}\label{Eqn:a-definition}
a_i = \frac{\overline{\rho u'_i}}{\overline{\rho}} = -\overline{u''_i}.
\end{equation}

\subsection{Averaged equations--Mean Flow}
Employing mass-weighted (Favre) averaging, ( we presume an ensemble-average), the Navier-Stokes equations become
\begin{eqnarray}\label{Eqn:NS_Avg}
\frac{\partial \overline{\rho} \tilde{u}_i}{\partial t} +\frac{\partial \overline{\rho} \tilde{u}_i \tilde{u}_n}{\partial x_n} 
+\frac{\partial R_{in}}{\partial x_n}
=  \frac{\partial \overline{\sigma}_{in}}{\partial x_n},
\end{eqnarray}
where
\begin{eqnarray} \label{Eqn:1ptRij}
R_{ij} = \overline{\rho u''_i u''_j}.
\end{eqnarray}
The continuity equation becomes
\begin{eqnarray}\label{Eqn:Continuity_Avg}
\frac{\partial \overline{\rho} }{\partial t} +\frac{\partial \overline{\rho} \tilde{u}_n}{\partial x_n} 
= 0,
\end{eqnarray}
and the averaged mass-fraction equations become
\begin{eqnarray}\label{Eqn:Concentration_Avg}
\frac{\partial \overline{\rho} \tilde{c}_k}{\partial t} +\frac{\partial \overline{\rho} \tilde{u}_n \tilde{c}_k}{\partial x_n}
-\frac{\partial \overline{\rho} a^k_n }{\partial x_n} 
= \frac{\partial}{\partial x_n}\left\{ {\cal D}\overline{\rho} \frac{\partial \tilde{c}_k}{\partial x_n}\right\}
+ \frac{\partial}{\partial x_n}\left\{ {\cal D}\overline{\rho \frac{\partial c''_k}{\partial x_n}}\right\},
\end{eqnarray}
where
\begin{eqnarray}
a^k_i = -\frac{\overline{\rho u''_i c''_k}}{\overline{\rho}}.
\end{eqnarray}
The averaged energy equation becomes
\begin{equation}\label{Eqn:Energy_Avg}
\begin{aligned}
\frac{\partial \overline{\rho} \tilde{I} }{\partial t} +\frac{\partial \overline{\rho} \tilde{u}_n \tilde{I}}{\partial x_n} 
+\frac{\partial S_n}{\partial x_n}
= \overline{\sigma}_{nm}\frac{\partial \overline{u}_n}{\partial x_m}
+\overline{\sigma'_{nm}\frac{\partial u'_n}{\partial x_m}} 
+ \frac{\partial }{\partial x_n}\left\{\kappa \frac{\partial \overline{T}}{\partial x_n}\right\},
 \end{aligned}
\end{equation}
where
\begin{eqnarray}
S_{i} = \overline{\rho u''_i I''}.
\end{eqnarray}
\subsection{Fluctuating Velocity and the Reynolds Stress Equation}
The exact, instantaneous equation for velocity is constructed from equations \ref{Eqn:NS} and \ref{Eqn:Continuity};
\begin{equation}\label{Eqn:NSVelocity}
\frac{\partial u_i}{\partial t} + u_n\frac{\partial u_i }{\partial x_n} = \frac{1}{\rho}\frac{\partial \sigma}{\partial x_i} .
\end{equation}
Likewise, the exact, ``averaged" equation for the velocity is constructed from equations  \ref{Eqn:NS_Avg} and \ref{Eqn:Continuity_Avg}
\begin{equation}\label{Eqn:NSVelocity_Avg}
\frac{\partial \tilde{u}_i}{\partial t} + \tilde{u}_n\frac{\partial \tilde{u}_i }{\partial x_n} 
+\frac{1}{\overline{\rho}}\frac{\partial R_{in}}{\partial x_n}=
+ \frac{1}{\overline{\rho}}\frac{\partial \overline{\sigma}_{in}}{\partial x_n}.
\end{equation}
Subtracting equation \ref{Eqn:NSVelocity_Avg} from equation \ref{Eqn:NSVelocity} yields the equation for the fluctuating velocity in the Favre-averaged sense;
\begin{equation}\label{Eqn:NSVelocity_Fluct1}
\begin{aligned}
\frac{\partial u''_i}{\partial t} + u_n\frac{\partial u''_i }{\partial x_n} 
+ u''_n\frac{\partial \tilde{u}_i }{\partial x_n} 
 -\frac{1}{\overline{\rho}}\frac{\partial R_{in}}{\partial x_n}
=
\left\{v\frac{\partial \sigma'_{in}}{\partial x_n} +\left(v- \frac{1}{\overline{\rho}}\right)\frac{\partial \overline{\sigma}_{in}}{\partial x_n}
\right\},
\end{aligned}
\end{equation}
where $v$ is the specific volume,
\begin{equation}
v =\frac{1}{\rho}.
\end{equation}
By employing equation \ref{Eqn:Continuity} in conjunction with equation \ref{Eqn:NSVelocity_Fluct1}, we can also construct
\begin{equation}\label{Eqn:NSVelocity_Fluct2}
\begin{aligned}
\frac{\partial \rho u''_i}{\partial t} + \frac{\partial \rho u_n u''_i }{\partial x_n} 
+ \rho u''_n\frac{\partial \tilde{u}_i }{\partial x_n} 
 -\frac{\rho}{\overline{\rho}}\frac{\partial R_{in}}{\partial x_n}
=\left\{\frac{\partial \sigma'_{in}}{\partial x_n} +\left(1- \frac{\rho}{\overline{\rho}}\right)\frac{\partial \overline{\sigma}_{in}}{\partial x_n}
\right\}.
\end{aligned}
\end{equation}
Using equations \ref{Eqn:NSVelocity_Fluct1} and \ref{Eqn:NSVelocity_Fluct2}, we can construct the ``unaveraged" Reynolds-Stress Transport equations,
\begin{equation}\label{Eqn:Reynolds_Unaveraged}
\begin{aligned}
\frac{\partial \rho u''_i u''_j}{\partial t} &+ \frac{\partial \rho u_n u''_i  u''_j}{\partial x_n} 
+ \rho u''_n u''_j\frac{\partial \tilde{u}_i }{\partial x_n} + \rho u''_n u''_i\frac{\partial \tilde{u}_j}{\partial x_n} 
 -\frac{\rho u''_j}{\overline{\rho}}\frac{\partial R_{in}}{\partial x_n}-\frac{\rho u''_i}{\overline{\rho}}\frac{\partial R_{jn}}{\partial x_n}\\
=& \left\{u''_j\frac{\partial \sigma'_{in}}{\partial x_n}+u''_i\frac{\partial \sigma'_{jn}}{\partial x_n}\right\}
+\left\{ 
\left(u''_j- \frac{\rho u''_j}{\overline{\rho}}\right)\frac{\partial \overline{\sigma}_{in}}{\partial x_n}
+\left(u''_i- \frac{\rho u''_i}{\overline{\rho}}\right)\frac{\partial \overline{\sigma}_{jn}}{\partial x_n}
\right\}.
\end{aligned}
\end{equation}
Averaging equation \ref{Eqn:Reynolds_Unaveraged} gives the exact Reynolds-stress transport equation;
\begin{equation}\label{Eqn:Reynolds1}
\begin{aligned}
\frac{\partial R_{ij}}{\partial t} + \frac{\partial \tilde{u}_n R_{ij}}{\partial x_n} + \frac{\partial R_{ijn}}{\partial x_n} 
&+ R_{jn}\frac{\partial \tilde{u}_i }{\partial x_n} + R_{in}\frac{\partial \tilde{u}_j}{\partial x_n} \\
=& \left\{\overline{u'_j\frac{\partial \sigma'_{in}}{\partial x_n}}+\overline{u'_i\frac{\partial \sigma'_{jn}}{\partial x_n}}\right\}
-\left\{ 
a_j\frac{\partial \overline{\sigma}_{in}}{\partial x_n}
+a_i\frac{\partial \overline{\sigma}_{jn}}{\partial x_n}
\right\}.
\end{aligned}
\end{equation}
where
\begin{equation}\label{Eqn:a_i}
a_i = -\overline{u''_i} = \frac{\overline{\rho' u'_i}}{\overline{\rho}}.
\end{equation}
\section{The Two-Point Reynolds Stress Equations}
Following Clark and Spitz, we define the two-point Reynolds Stress tensor as
\begin{equation}\label{Eqn:2ptRij}
\begin{aligned} 
R_{ij}\left({\bf x_1},{\bf x_2}\right) 
&= \frac{1}{2}\left[\overline{
  \rho\left({\bf x_1}\right) u''_i\left({\bf x_1}\right) u''_j\left({\bf x_2}\right)
+\rho\left({\bf x_2}\right) u''_j\left({\bf x_2}\right) u''_i\left({\bf x_1}\right)}\right] \\
&= \frac{1}{2}\overline{
  \left[\rho\left({\bf x_1}\right) +\rho\left({\bf x_2}\right) \right]  u''_i\left({\bf x_1}\right) u''_j\left({\bf x_2}\right)
}.
\end{aligned}
\end{equation}
This definition of the Reynolds stress tensor satisfies the symmetry property
$$R_{ij}\left({\bf x_1},{\bf x_2}\right) =R_{ji}\left({\bf x_2},{\bf x_1}\right), $$
and is equal to the one-point definition, equation \ref{Eqn:1ptRij}, for the case of ${\bf x_1} = {\bf x_2} = {\bf x}$.

The evolution equation for the two-point Reynolds Stress tensor can be constructed from equations \ref{Eqn:NSVelocity_Fluct1} and \ref{Eqn:NSVelocity_Fluct2}.  First we derive
\begin{equation}\label{Eqn:2ptRijEquation1}
\begin{aligned} 
&\frac{\partial \rho\left({\bf x_1}\right)u''_i\left({\bf x_1}\right)u''_j\left({\bf x_2}\right)}{\partial t}
+\frac{\partial  \rho\left({\bf x_1}\right)u''_i\left({\bf x_1}\right)u''_j\left({\bf x_2}\right)u_n\left({\bf x_1}\right)}{\partial x_{1n}}\\
&+\frac{\partial  \rho\left({\bf x_1}\right)u''_i\left({\bf x_1}\right)u''_j\left({\bf x_2}\right)u_n\left({\bf x_2}\right)}{\partial x_{2n}}
- \rho\left({\bf x_1}\right)u''_i\left({\bf x_1}\right)u''_j\left({\bf x_2}\right)\frac{\partial u_n\left({\bf x_2}\right)}{\partial x_{2n}}\\
&+\rho\left({\bf x_1}\right)u''_i\left({\bf x_1}\right)u''_n\left({\bf x_2}\right)\frac{\partial \tilde{u}_j\left({\bf x_2}\right)}{\partial x_{2n}}
+\rho\left({\bf x_1}\right)u''_n\left({\bf x_1}\right)u''_j\left({\bf x_2}\right)\frac{\partial \tilde{u}_i\left({\bf x_1}\right)}{\partial x_{1n}}\\
&-\frac{\rho\left({\bf x_1}\right) u''_i\left({\bf x_1}\right) }{ \overline{\rho}\left({\bf x_2}\right)} 
\frac{\partial R_{jn}\left({\bf x_2},{\bf x_2}\right)}{\partial x_{2n}}
-\frac{\rho\left({\bf x_1}\right) u''_j\left({\bf x_2}\right) }{ \overline{\rho}\left({\bf x_1}\right)} 
\frac{\partial R_{in}\left({\bf x_1},{\bf x_1}\right)}{\partial x_{1n}}\\
&=\rho\left({\bf x_1}\right)u''_i\left({\bf x_1}\right)v\left({\bf x_2}\right)\frac{\partial \sigma'_{jn}\left({\bf x_2}\right)}{\partial x_{2n}}
+u''_j\left({\bf x_2}\right)\frac{\partial \sigma'_{in}\left({\bf x_1}\right)}{\partial x_{1n}}\\
&+\left[\rho\left({\bf x_1}\right)v\left({\bf x_2}\right)u''_i\left({\bf x_1}\right)-\frac{\rho\left({\bf x_1}\right)u''_i\left({\bf x_1}\right)}{\overline{\rho}\left({\bf x_2}\right)}\right]\frac{\partial \overline{\sigma}_{jn}\left({\bf x_2}\right)}{\partial x_{2n}}\\
&+\left[u''_j\left({\bf x_2}\right)-\frac{\rho\left({\bf x_1}\right)u''_j\left({\bf x_2}\right)}{\overline{\rho}\left({\bf x_1}\right)}\right]\frac{\partial \overline{\sigma}_{in}\left({\bf x_1}\right)}{\partial x_{1n}}
\end{aligned}
\end{equation}
Next, interchange ${\bf x_1}$ and ${\bf x_2}$ and the indices $i$ and $j$;
\begin{equation}\label{Eqn:2ptRijEquation2}
\begin{aligned} 
&\frac{\partial \rho\left({\bf x_2}\right)u''_j\left({\bf x_2}\right)u''_i\left({\bf x_1}\right)}{\partial t}
+\frac{\partial  \rho\left({\bf x_2}\right)u''_j\left({\bf x_2}\right)u''_i\left({\bf x_1}\right)u_n\left({\bf x_2}\right)}{\partial x_{2n}}\\
&+\frac{\partial  \rho\left({\bf x_2}\right)u''_j\left({\bf x_2}\right)u''_i\left({\bf x_1}\right)u_n\left({\bf x_1}\right)}{\partial x_{1n}}
- \rho\left({\bf x_2}\right)u''_j\left({\bf x_2}\right)u''_i\left({\bf x_1}\right)\frac{\partial u_n\left({\bf x_1}\right)}{\partial x_{1n}}\\
&+\rho\left({\bf x_2}\right)u''_j\left({\bf x_2}\right)u''_n\left({\bf x_1}\right)\frac{\partial \tilde{u}_i\left({\bf x_1}\right)}{\partial x_{1n}}
+\rho\left({\bf x_2}\right)u''_n\left({\bf x_2}\right)u''_i\left({\bf x_1}\right)\frac{\partial \tilde{u}_j\left({\bf x_2}\right)}{\partial x_{2n}}\\
&-\frac{\rho\left({\bf x_2}\right) u''_j\left({\bf x_2}\right) }{ \overline{\rho}\left({\bf x_1}\right)} 
\frac{\partial R_{in}\left({\bf x_1},{\bf x_1}\right)}{\partial x_{1n}}
-\frac{\rho\left({\bf x_2}\right) u''_i\left({\bf x_1}\right) }{ \overline{\rho}\left({\bf x_2}\right)} 
\frac{\partial R_{jn}\left({\bf x_2},{\bf x_2}\right)}{\partial x_{2n}}\\
&= \rho\left({\bf x_2}\right)u''_j\left({\bf x_2}\right)v\left({\bf x_1}\right)\frac{\partial \sigma'_{in}\left({\bf x_1}\right)}{\partial x_{1n}}
+u''_i\left({\bf x_1}\right)\frac{\partial \sigma'_{jn}\left({\bf x_2}\right)}{\partial x_{2n}}\\
&+\left[\rho\left({\bf x_2}\right)v\left({\bf x_1}\right)u''_j\left({\bf x_2}\right)-\frac{\rho\left({\bf x_2}\right)u''_j\left({\bf x_2}\right)}{\overline{\rho}\left({\bf x_1}\right)}\right]\frac{\partial \overline{\sigma}_{in}\left({\bf x_1}\right)}{\partial x_{1n}}\\
&+\left[u''_i\left({\bf x_1}\right)-\frac{\rho\left({\bf x_2}\right)u''_i\left({\bf x_1}\right)}{\overline{\rho}\left({\bf x_2}\right)}\right]\frac{\partial \overline{\sigma}_{jn}\left({\bf x_2}\right)}{\partial x_{2n}}
\end{aligned}
\end{equation}
Next, we add equation \ref{Eqn:2ptRijEquation1} to equation \ref{Eqn:2ptRijEquation2} and divide the result by 2;
\begin{equation}\label{Eqn:2ptRijEquation3}
\begin{aligned} 
&\frac{\partial}{\partial t}\left\{\frac{\left[\rho\left({\bf x_1}\right)+\rho\left({\bf x_2}\right)\right]}{2}u''_i\left({\bf x_1}\right)u''_j\left({\bf x_2}\right)\right\}\\
&+\frac{\partial }{\partial x_{1n}}\left\{\frac{\left[\rho\left({\bf x_1}\right)+\rho\left({\bf x_2}\right)\right]}{2}u''_i\left({\bf x_1}\right)u''_j\left({\bf x_2}\right)u_n\left({\bf x_1}\right)\right\}\\
&+\frac{\partial }{\partial x_{2n}}\left\{\frac{\left[\rho\left({\bf x_1}\right)+\rho\left({\bf x_2}\right)\right]}{2}u''_i\left({\bf x_1}\right)u''_j\left({\bf x_2}\right)u_n\left({\bf x_2}\right)\right\}\\
&-\frac{1}{2} \left\{\rho\left({\bf x_1}\right)u''_i\left({\bf x_1}\right)u''_j\left({\bf x_2}\right)\frac{\partial u_n\left({\bf x_2}\right)}{\partial x_{2n}}
  +\rho\left({\bf x_2}\right)u''_i\left({\bf x_1}\right)u''_j\left({\bf x_2}\right)\frac{\partial u_n\left({\bf x_1}\right)}{\partial x_{1n}}\right\}\\
&+\left\{\frac{\left[\rho\left({\bf x_1}\right) + \rho\left({\bf x_2}\right)\right]}{2}u''_n\left({\bf x_1}\right)u''_j\left({\bf x_2}\right)\right\}\frac{\partial \tilde{u}_i\left({\bf x_1}\right)}{\partial x_{1n}}\\
&+\left\{\frac{\left[\rho\left({\bf x_1}\right) + \rho\left({\bf x_2}\right)\right]}{2}u''_i\left({\bf x_1}\right)u''_n\left({\bf x_2}\right)\right\}\frac{\partial \tilde{u}_j\left({\bf x_2}\right)}{\partial x_{2n}}\\
&-\frac{\left[\rho\left({\bf x_1}\right) + \rho\left({\bf x_2}\right)\right]}{2}\frac{u''_i\left({\bf x_1}\right)}{\overline{\rho}\left({\bf x_2}\right)}\frac{\partial R_{jn}\left({\bf x_2},{\bf x_2}\right)}{\partial x_{2n}}\\
&-\frac{\left[\rho\left({\bf x_1}\right) + \rho\left({\bf x_2}\right)\right]}{2}\frac{u''_j\left({\bf x_2}\right)}{\overline{\rho}\left({\bf x_1}\right)}\frac{\partial R_{in}\left({\bf x_1},{\bf x_1}\right)}{\partial x_{1n}} \\ 
&=\frac{\left[\rho\left({\bf x_1}\right) + \rho\left({\bf x_2}\right)\right]}{2}
u''_i\left({\bf x_1}\right) v\left({\bf x_2}\right)\frac{\partial \sigma'_{jn}\left({\bf x_2}\right)}{\partial x_{2n}}\\
&+\frac{\left[\rho\left({\bf x_1}\right) + \rho\left({\bf x_2}\right)\right]}{2}
u''_j\left({\bf x_2}\right) v\left({\bf x_1}\right)\frac{\partial \sigma'_{in}\left({\bf x_1}\right)}{\partial x_{1n}}\\
&+\frac{\left[\rho\left({\bf x_1}\right) + \rho\left({\bf x_2}\right)\right]}{2}\left\{
u''_i\left({\bf x_1}\right)\left[v\left({\bf x_2}\right) -\frac{1}{\overline{\rho}\left({\bf x_2}\right)}\right]\right\}\frac{\partial \overline{\sigma}_{jn}\left({\bf x_2}\right)}{\partial x_{2n}}\\
&+\frac{\left[\rho\left({\bf x_1}\right) + \rho\left({\bf x_2}\right)\right]}{2}\left\{
u''_j\left({\bf x_2}\right)\left[v\left({\bf x_1}\right) -\frac{1}{\overline{\rho}\left({\bf x_1}\right)}\right]\right\}\frac{\partial \overline{\sigma}_{in}\left({\bf x_1}\right)}{\partial x_{1n}}
\end{aligned}
\end{equation}

Averaging this equation yields
\begin{equation}\label{Eqn:2ptRijEquation}
\begin{aligned} 
&\frac{\partial R_{ij}\left({\bf x_1},{\bf x_2}\right)}{\partial t}
+\frac{\partial }{\partial x_{1n}}\left\{R_{ij}\left({\bf x_1},{\bf x_2}\right)\tilde{u}_n\left({\bf x_1}\right)\right\}
+\frac{\partial }{\partial x_{2n}}\left\{R_{ij}\left({\bf x_1},{\bf x_2}\right)\tilde{u}_n\left({\bf x_2}\right)\right\}\\
&+\frac{\partial }{\partial x_{1n}}\left\{R_{ijn}\left({\bf x_1},{\bf x_2},{\bf x_1}\right)\right\}
+\frac{\partial }{\partial x_{2n}}\left\{R_{ijn}\left({\bf x_1},{\bf x_2},{\bf x_2}\right)\right\}\\
&-\frac{1}{2} \left\{T_{ij}\left({\bf x_1},{\bf x_2}\right)\frac{\partial u_n\left({\bf x_2}\right)}{\partial x_{2n}}
  +T_{ji}\left({\bf x_2},{\bf x_1}\right)\frac{\partial u_n\left({\bf x_1}\right)}{\partial x_{1n}}\right\}\\
&+R_{nj}\left({\bf x_1},{\bf x_2}\right)\frac{\partial \tilde{u}_i\left({\bf x_1}\right)}{\partial x_{1n}}
+R_{in}\left({\bf x_1},{\bf x_2}\right)\frac{\partial \tilde{u}_j\left({\bf x_2}\right)}{\partial x_{2n}}\\
&-\frac{Q_i\left({\bf x_1},{\bf x_2}\right)}{\overline{\rho}\left({\bf x_2}\right)}\frac{\partial R_{jn}\left({\bf x_2},{\bf x_2}\right)}{\partial x_{2n}}
-\frac{Q_j\left({\bf x_2},{\bf x_1}\right)}{\overline{\rho}\left({\bf x_1}\right)}\frac{\partial R_{in}\left({\bf x_1},{\bf x_1}\right)}{\partial x_{1n}} \\ 
&= a_{i}\left({\bf x_1},{\bf x_2}\right)\frac{\partial \overline{\sigma}_{jn}\left({\bf x_2}\right)}{\partial x_{2n}}
+a_{j}\left({\bf x_2},{\bf x_1}\right)\frac{\partial \overline{\sigma}_{in}\left({\bf x_1}\right)}{\partial x_{1n}}\\
&+\Omega_{ij}\left({\bf x_1},{\bf x_2}\right)
 +\Omega_{ji}\left({\bf x_2},{\bf x_1}\right)
\end{aligned}
\end{equation}
where
\begin{equation}
R_{ij}\left({\bf x_1},{\bf x_2}\right) = \frac{1}{2}\overline{\left[\rho\left({\bf x_1}\right)+\rho\left({\bf x_2}\right)\right]u''_i\left({\bf x_1}\right)u''_j\left({\bf x_2}\right)}
\end{equation}
\begin{equation}
R_{ijk}\left({\bf x_1},{\bf x_2},{\bf x_3}\right) = \frac{1}{2}\overline{\left[\rho\left({\bf x_1}\right)+\rho\left({\bf x_2}\right)\right]u''_i\left({\bf x_1}\right)u''_j\left({\bf x_2}\right)u''_k\left({\bf x_3}\right)}
\end{equation}
\begin{equation}
T_{ij}\left({\bf x_1},{\bf x_2}\right) = \overline{\rho\left({\bf x_1}\right)u''_i\left({\bf x_1}\right)u''_j\left({\bf x_2}\right)}
\end{equation}
\begin{equation}
Q_{i}\left({\bf x_1},{\bf x_2}\right) =  \frac{1}{2}\overline{\left[\rho\left({\bf x_1}\right)+\rho\left({\bf x_2}\right)\right]u''_i\left({\bf x_1}\right)}
=  \frac{1}{2}\overline{\rho\left({\bf x_2}\right)u''_i\left({\bf x_1}\right)}
\end{equation}
and
\begin{equation}
\begin{aligned}
a_{i}\left({\bf x_1},{\bf x_2}\right) &= -\overline{\frac{\left[\rho\left({\bf x_1}\right) + \rho\left({\bf x_2}\right)\right]}{2}\left\{
u''_i\left({\bf x_1}\right)\left[v\left({\bf x_2}\right) -\frac{1}{\overline{\rho}\left({\bf x_2}\right)}\right]\right\}}\\
&= -\frac{1}{2}\left\{
\overline{\rho\left({\bf x_1}\right)u''_i\left({\bf x_1}\right)v\left({\bf x_2}\right)}
+\overline{u''_i\left({\bf x_1}\right)}
-\frac{\overline{ \rho\left({\bf x_2}\right)u''_i\left({\bf x_1}\right)}}{\overline{\rho}\left({\bf x_2}\right)}
\right\}.
\end{aligned}
\end{equation}
Note that
\begin{equation}
\begin{aligned}
\Omega_{ij}\left({\bf x_1},{\bf x_2}\right) = -\Pi_{ij}\left({\bf x_1},{\bf x_2}\right) - {\cal E}_{ij}\left({\bf x_1},{\bf x_2}\right)
\end{aligned}
\end{equation}
where
\begin{equation}
\begin{aligned}
\Pi_{ij}\left({\bf x_1},{\bf x_2}\right)&=\overline{\frac{\left[\rho\left({\bf x_1}\right) + \rho\left({\bf x_2}\right)\right]}{2}
u''_i\left({\bf x_1}\right) v\left({\bf x_2}\right)\frac{\partial p'\left({\bf x_2}\right)}{\partial x_{2j}}}\\
&=\frac{1}{2}\left\{
\overline{\left[1+\rho\left({\bf x_1}\right)v\left({\bf x_2}\right)\right]u''_i\left({\bf x_1}\right)
\frac{\partial p'\left({\bf x_2}\right)}{\partial x_{2j}} 
}
\right\}
\end{aligned}
\end{equation}
\begin{equation}
\begin{aligned}
{\cal E}_{ij} \left({\bf x_1},{\bf x_2}\right) &= -\overline{\frac{\left[\rho\left({\bf x_1}\right) + \rho\left({\bf x_2}\right)\right]}{2}
u''_i\left({\bf x_1}\right) v\left({\bf x_2}\right)\frac{\partial \tau'_{jn}\left({\bf x_2}\right)}{\partial x_{2n}}}\\
&= -\frac{1}{2}\left\{
\overline{\left[1 + \rho\left({\bf x_1}\right)v\left({\bf x_2}\right)\right]
u''_i\left({\bf x_1}\right) \frac{\partial \tau'_{jn}\left({\bf x_2}\right)}{\partial x_{2n}}}
\right\}\end{aligned}
\end{equation}

\subsection{The one-point limit}
Note that in the one-point limit, (${\bf x_1} = {\bf x_2} = {\bf x}$, these correlations become
\begin{equation}
R_{ij}\left({\bf x},{\bf x}\right) = \overline{\rho\left({\bf x}\right)u''_i\left({\bf x}\right)u''_j\left({\bf x}\right)} = R_{ij}\left({\bf x}\right),
\end{equation}
\begin{equation}
R_{ijk}\left({\bf x},{\bf x},{\bf x}\right) = \overline{\rho\left({\bf x}\right)u''_i\left({\bf x}\right)u''_j\left({\bf x}\right)u''_k\left({\bf x}\right)} = R_{ijk}\left({\bf x}\right),
\end{equation}
\begin{equation}
T_{ij}\left({\bf x},{\bf x}\right) = \overline{\rho\left({\bf x}\right)u''_i\left({\bf x}\right)u''_j\left({\bf x}\right)} = R_{ij}\left({\bf x}\right),
\end{equation}
\begin{equation}
Q_{i}\left({\bf x},{\bf x}\right) =  \frac{1}{2}\overline{\rho\left({\bf x}\right)u''_i\left({\bf x}\right)} = 0,
\end{equation}
\begin{equation}
\begin{aligned}
a_{i}\left({\bf x},{\bf x}\right) &= -\frac{1}{2}\left\{
\overline{\rho\left({\bf x}\right)u''_i\left({\bf x}\right)v\left({\bf x}\right)}
+\overline{u''_i\left({\bf x}\right)}
-\frac{\overline{ \rho\left({\bf x}\right)u''_i\left({\bf x}\right)}}{\overline{\rho}\left({\bf x}\right)}\right\}\\
 &= -\overline{u''_i\left({\bf x}\right)} = a_i\left({\bf x}\right),
\end{aligned}
\end{equation}

\begin{equation}
\begin{aligned}
\Pi_{ij}\left({\bf x},{\bf x}\right)&=\frac{1}{2}\left\{
\overline{
\left[1+\rho\left({\bf x}\right)v\left({\bf x}\right)\right]u''_i\left({\bf x}\right)
\frac{\partial p'\left({\bf x}\right)}{\partial x_{j}} 
} \right\}
= \overline{
u''_i\left({\bf x}\right)
\frac{\partial p'\left({\bf x}\right)}{\partial x_{j}} 
}
\end{aligned}
\end{equation}

\begin{equation}
\begin{aligned}
{\cal E}_{ij} \left({\bf x},{\bf x}\right) &= -\frac{1}{2}\left\{
\overline{\left[1 + \rho\left({\bf x}\right)v\left({\bf x}\right)\right]
u''_i\left({\bf x}\right) \frac{\partial \tau'_{jn}\left({\bf x}\right)}{\partial x_{n}}}
\right\}
= -
\overline{
u''_i\left({\bf x}\right) \frac{\partial \tau'_{jn}\left({\bf x}\right)}{\partial x_{n}}}
\end{aligned}
\end{equation}
\section{The $a_i$ equation}
We first note that
\begin{equation} \label{Eqn:RhoCentered}
\begin{aligned}
\frac{\partial}{\partial t}\left\{\frac{\rho\left({\bf x_1}\right) + \rho\left({\bf x_2}\right)}{2}\right\}
+
\frac{1}{2}\left\{
\frac{\partial \rho\left({\bf x_1}\right) u_n\left({\bf x_1}\right)}{\partial x_{1n}}
+\frac{\partial \rho\left({\bf x_2}\right) u_n\left({\bf x_2}\right)}{\partial x_{2n}}
\right\}=0,
\end{aligned}
\end{equation}
\begin{equation}
\begin{aligned}
\frac{\partial v\left({\bf x}\right)}{\partial t}
+ \frac{\partial v\left({\bf x}\right) u_n\left({\bf x}\right)}{\partial x_n} = 2 v\left({\bf x}\right)\frac{\partial  u_n\left({\bf x}\right)}{\partial x_n}
\end{aligned}
\end{equation}
and
\begin{equation}
\begin{aligned}
\frac{\partial }{\partial t} \left\{
\frac{1}{\overline{\rho}\left({\bf x}\right)}
\right\}
+ \frac{\partial }{\partial x_n}
\left\{
\frac{ \tilde{u}_n\left({\bf x}\right) }{ \overline{\rho}\left({\bf x}\right)}
\right\}
=\frac{2}{\overline{\rho}\left({\bf x}\right)}\frac{\partial \tilde{u}_n\left({\bf x}\right)}{\partial x_n},
\end{aligned}
\end{equation}
so that
\begin{equation}
\begin{aligned}
\frac{\partial }{\partial t} \left\{v\left({\bf x}\right) - 
\frac{1}{\overline{\rho}\left({\bf x}\right)}
\right\}
+ \frac{\partial }{\partial x_n}
\left\{v\left({\bf x}\right) u_n\left({\bf x}\right) -
\frac{ \tilde{u}_n\left({\bf x}\right) }{ \overline{\rho}\left({\bf x}\right)}
\right\}
=2\left\{ v\left({\bf x}\right)\frac{\partial  u_n\left({\bf x}\right)}{\partial x_n}
-\frac{1}{\overline{\rho}\left({\bf x}\right)}\frac{\partial \tilde{u}_n\left({\bf x}\right)}{\partial x_n}
\right\},
\end{aligned}
\end{equation}
or equivalently,
\begin{equation}
\begin{aligned}
\frac{\partial }{\partial t} \left\{v\left({\bf x}\right) - 
\frac{1}{\overline{\rho}\left({\bf x}\right)}
\right\}
&+ \frac{\partial }{\partial x_n}
\left\{\tilde{u}_n\left({\bf x}\right
)\left[v\left({\bf x}\right)  -
\frac{ 1 }{ \overline{\rho}\left({\bf x}\right)}\right]
\right\}
+ \frac{\partial v\left({\bf x}\right)u''_n\left({\bf x}\right)}{\partial x_n}
\\
&=2\left\{ \left[v\left({\bf x}\right)
-\frac{1}{\overline{\rho}\left({\bf x}\right)}\right]\frac{\partial \tilde{u}_n\left({\bf x}\right)}{\partial x_n}\right\}
+2v\left({\bf x}\right)\frac{\partial  u''_n\left({\bf x}\right)}{\partial x_n},
\end{aligned}
\end{equation}
or equivalently,
\begin{equation} \label{Eqn:PhiEvo}
\begin{aligned}
\frac{\partial \phi\left({\bf x}\right)}{\partial t} &+ \frac{\partial \tilde{u}_n\left({\bf x}\right
)\phi\left({\bf x}\right)}{\partial x_n}
+ \frac{\partial v\left({\bf x}\right)u''_n\left({\bf x}\right)}{\partial x_n}
\\
&=2\left\{\phi\left({\bf x}\right)\frac{\partial \tilde{u}_n\left({\bf x}\right)}{\partial x_n}\right\}
+2v\left({\bf x}\right)\frac{\partial  u''_n\left({\bf x}\right)}{\partial x_n},
\end{aligned}
\end{equation}
where we have let
\begin{equation}\label{Eqn:PhiDefNew}
\phi\left({\bf x}\right) = \left[v\left({\bf x}\right)
-\frac{1}{\overline{\rho}\left({\bf x}\right)}\right]
\end{equation}
Next note that 
\begin{equation} \label{Eqn:RhoCenteredPhi}
\begin{aligned}
\frac{\partial}{\partial t}&\left\{\left[\frac{\rho\left({\bf x_1}\right) + \rho\left({\bf x_2}\right)}{2}\right]\phi\left({\bf x_2}\right)\right\}
+
\frac{1}{2}\left\{
\frac{\partial \rho\left({\bf x_1}\right) u_n\left({\bf x_1}\right)}{\partial x_{1n}}
+\frac{\partial \rho\left({\bf x_2}\right) u_n\left({\bf x_2}\right)}{\partial x_{2n}}
\right\}\phi\left({\bf x_2}\right)\\
+&\left[\frac{\rho\left({\bf x_1}\right) + \rho\left({\bf x_2}\right)}{2}\right]\left\{
\frac{\partial \tilde{u}_n\left({\bf x_2}\right
)\phi\left({\bf x_2}\right)}{\partial x_{2n}}
+ \frac{\partial v\left({\bf x_2}\right)u''_n\left({\bf x_2}\right)}{\partial x_{2n}}
\right\}\\
=&2\left\{\left[\frac{\rho\left({\bf x_1}\right) + \rho\left({\bf x_2}\right)}{2}\right]\phi\left({\bf x_2}\right)\frac{\partial \tilde{u}_n\left({\bf x_2}\right)}{\partial x_{2n}}\right\}
+2\left[\frac{\rho\left({\bf x_1}\right) + \rho\left({\bf x_2}\right)}{2}\right]v\left({\bf x_2}\right)\frac{\partial  u''_n\left({\bf x_2}\right)}{\partial x_{2n}},
\end{aligned}
\end{equation},
or equivalently,
\begin{equation} \label{Eqn:RhoCenteredPhi2}
\begin{aligned}
\frac{\partial}{\partial t}&\left\{\rho_c\left({\bf x_1},{\bf x_2}\right)\phi\left({\bf x_2}\right)\right\}
+
\frac{1}{2}\left\{
\frac{\partial \rho\left({\bf x_1}\right) u_n\left({\bf x_1}\right)}{\partial x_{1n}}
+\frac{\partial \rho\left({\bf x_2}\right) u_n\left({\bf x_2}\right)}{\partial x_{2n}}
\right\}\phi\left({\bf x_2}\right)\\
+&\rho_c\left({\bf x_1},{\bf x_2}\right)\left\{
\frac{\partial \tilde{u}_n\left({\bf x_2}\right)\phi\left({\bf x_2}\right)}{\partial x_{2n}}
+ \frac{\partial v\left({\bf x_2}\right)u''_n\left({\bf x_2}\right)}{\partial x_{2n}}
\right\}\\
=&2\left\{\rho_c\left({\bf x_1},{\bf x_2}\right)\phi\left({\bf x_2}\right)\frac{\partial \tilde{u}_n\left({\bf x_2}\right)}{\partial x_{2n}}\right\}
+2\rho_c\left({\bf x_1},{\bf x_2}\right)v\left({\bf x_2}\right)\frac{\partial  u''_n\left({\bf x_2}\right)}{\partial x_{2n}},
\end{aligned}
\end{equation}
where
\begin{equation}\rho_c\left({\bf x_1},{\bf x_2}\right) = \left[\frac{\rho\left({\bf x_1}\right) + \rho\left({\bf x_2}\right)}{2}\right].
\end{equation}
Using equation \ref{Eqn:NSVelocity_Fluct1} along with equation \ref{Eqn:RhoCenteredPhi2} allows us to construct the exact transport equation for $a_i\left({\bf x_1},{\bf x_2}\right)$;
\begin{equation}\label{Eqn:a_tranport4}
\begin{aligned}
\frac{\partial \rho_c\left({\bf x_1},{\bf x_2}\right)u''_i\left({\bf x_1}\right)\phi\left({\bf x_2}\right)}{\partial t} 
&+\frac{\partial  \rho_c\left({\bf x_1},{\bf x_2}\right)u''_i \left({\bf x_1}\right)\phi\left({\bf x_2}\right)u_n\left({\bf x_1}\right)}{\partial x_{1n}} \\
&+ \rho_c\left({\bf x_1},{\bf x_2}\right)\phi\left({\bf x_2}\right)u''_n\left({\bf x_1}\right)\frac{\partial \tilde{u}_i\left({\bf x_1}\right) }{\partial x_{1n}} \\
 &-\frac{\rho_c\left({\bf x_1},{\bf x_2}\right)\phi\left({\bf x_2}\right)}{\overline{\rho}\left({\bf x_1}\right)}\frac{\partial R_{in}\left({\bf x_1},{\bf x_1}\right)}{\partial x_{1n}}\\
&+\frac{1}{2}\phi\left({\bf x_2}\right)u''_i\left({\bf x_1}\right)\left\{
\frac{\partial \rho\left({\bf x_2}\right) u_n\left({\bf x_2}\right)}{\partial x_{2n}}
-\frac{\partial \rho\left({\bf x_2}\right) u_n\left({\bf x_1}\right)}{\partial x_{1n}}
\right\}\\
+&\rho_c\left({\bf x_1},{\bf x_2}\right)u''_i\left({\bf x_1}\right)\left\{
\frac{\partial \tilde{u}_n\left({\bf x_2}\right)\phi\left({\bf x_2}\right)}{\partial x_{2n}}
+ \frac{\partial v\left({\bf x_2}\right)u''_n\left({\bf x_2}\right)}{\partial x_{2n}}
\right\}\\
=&\rho_c\left({\bf x_1},{\bf x_2}\right)\phi\left({\bf x_2}\right)v\left({\bf x_1}\right)\frac{\partial \sigma'_{in}\left({\bf x_1}\right)}{\partial x_{1n}} \\
&+\rho_c\left({\bf x_1},{\bf x_2}\right)\phi\left({\bf x_2}\right)\phi\left({\bf x_1}\right)\frac{\partial \overline{\sigma}_{in}\left({\bf x_1}\right)}{\partial x_{1n}}\\
+&2\left\{\rho_c\left({\bf x_1},{\bf x_2}\right)u''_i\left({\bf x_1}\right)\phi\left({\bf x_2}\right)\frac{\partial \tilde{u}_n\left({\bf x_2}\right)}{\partial x_{2n}}\right\}\\
+&2\rho_c\left({\bf x_1},{\bf x_2}\right)u''_i\left({\bf x_1}\right)v\left({\bf x_2}\right)\frac{\partial  u''_n\left({\bf x_2}\right)}{\partial x_{2n}},
\end{aligned}
\end{equation}

Averaging gives
\begin{equation}\label{Eqn:a_tranport5}
\begin{aligned}
\frac{\partial a_i\left({\bf x_1},{\bf x_2}\right)}{\partial t} 
&+\frac{\partial  a_i\left({\bf x_1},{\bf x_2}\right)\tilde{u}_n\left({\bf x_1}\right)}{\partial x_{1n}}
+\frac{\partial  \overline{\rho_c\left({\bf x_1},{\bf x_2}\right)u''_i \left({\bf x_1}\right)\phi\left({\bf x_2}\right)u''_n\left({\bf x_1}\right)}}{\partial x_{1n}} \\
&+ a_i\left({\bf x_1},{\bf x_2}\right)\frac{\partial \tilde{u}_i\left({\bf x_1}\right) }{\partial x_{1n}} \\
 &-\frac{\rho_c\left({\bf x_1},{\bf x_2}\right)\phi\left({\bf x_2}\right)}{\overline{\rho}\left({\bf x_1}\right)}\frac{\partial R_{in}\left({\bf x_1},{\bf x_1}\right)}{\partial x_{1n}}\\
&+\frac{1}{2}\phi\left({\bf x_2}\right)u''_i\left({\bf x_1}\right)\left\{
\frac{\partial \rho\left({\bf x_2}\right) u_n\left({\bf x_2}\right)}{\partial x_{2n}}
-\frac{\partial \rho\left({\bf x_2}\right) u_n\left({\bf x_1}\right)}{\partial x_{1n}}
\right\}\\
+&\rho_c\left({\bf x_1},{\bf x_2}\right)u''_i\left({\bf x_1}\right)\left\{
\frac{\partial \tilde{u}_n\left({\bf x_2}\right)\phi\left({\bf x_2}\right)}{\partial x_{2n}}
+ \frac{\partial v\left({\bf x_2}\right)u''_n\left({\bf x_2}\right)}{\partial x_{2n}}
\right\}\\
=&\rho_c\left({\bf x_1},{\bf x_2}\right)\phi\left({\bf x_2}\right)v\left({\bf x_1}\right)\frac{\partial \sigma'_{in}\left({\bf x_1}\right)}{\partial x_{1n}} \\
&+\beta\left({\bf x_1},{\bf x_2}\right)\frac{\partial \overline{\sigma}_{in}\left({\bf x_1}\right)}{\partial x_{1n}}\\
+&2\left\{a_i\left({\bf x_1},{\bf x_2}\right)\frac{\partial \tilde{u}_n\left({\bf x_2}\right)}{\partial x_{2n}}\right\}\\
+&2\rho_c\left({\bf x_1},{\bf x_2}\right)u''_i\left({\bf x_1}\right)v\left({\bf x_2}\right)\frac{\partial  u''_n\left({\bf x_2}\right)}{\partial x_{2n}},
\end{aligned}
\end{equation}
The two-point quantity corresponding to the BHR ``b" variable is
\begin{eqnarray}\label{Eqn:b_2pt1}
\beta\left({\bf x_1},{\bf x_2}\right)
= \overline{\rho_c\left({\bf x_1},{\bf x_2}\right)\phi\left({\bf x_1}\right)\phi\left({\bf x_2}\right) }
\end{eqnarray}
Note that
\begin{equation}\label{Eqn:Beta_2pt}
\begin{aligned}
\overline{\rho_c\left({\bf x_1},{\bf x_2}\right)\phi\left({\bf x_1}\right)\phi\left({\bf x_2}\right)}
&= \frac{1}{2}\overline{\left[
\rho\left({\bf x_1}\right) \phi\left({\bf x_1}\right) \phi\left({\bf x_2}\right)+
\rho\left({\bf x_2}\right) \phi\left({\bf x_2}\right) \phi\left({\bf x_1}\right) 
\right]} \\
& =  \frac{1}{2}\overline{\left\{
\left[ 
1-\frac{\rho\left({\bf x_1}\right)}{\overline{\rho}\left({\bf x_1}\right)}
\right]\phi\left({\bf x_2}\right)+
 \left[
1-\frac{\rho\left({\bf x_2}\right)}{\overline{\rho}\left({\bf x_2}\right)}
\right]\phi\left({\bf x_1}\right)
\right\} } \\
& =  -\frac{1}{2} \overline{ \left\{
\left[ 
\frac{\rho'\left({\bf x_1}\right)}{\overline{\rho}\left({\bf x_1}\right)}
\right]\phi\left({\bf x_2}\right)+
 \left[
\frac{\rho'\left({\bf x_2}\right)}{\overline{\rho}\left({\bf x_2}\right)}
\right]\phi\left({\bf x_1}\right)
\right\} } \\
& =  -\frac{1}{2} \overline{ \left\{
\left[ 
\frac{\rho'\left({\bf x_1}\right)v'\left({\bf x_2}\right)}{\overline{\rho}\left({\bf x_1}\right)}
\right]+
 \left[
\frac{\rho'\left({\bf x_2}\right)v'\left({\bf x_1}\right)}{\overline{\rho}\left({\bf x_2}\right)}
\right]
\right\} }
\end{aligned}
\end{equation}
In the one-point limit this becomes
\begin{equation}\label{Eqn:b_1pt}
\begin{aligned}
\overline{\rho_c\left({\bf x},{\bf x}\right)\phi\left({\bf x}\right)\phi\left({\bf x}\right) } 
&= - 
\left[ 
\overline{ 
\frac{\rho'\left({\bf x}\right)v'\left({\bf x}\right)}{\overline{\rho}\left({\bf x}\right)}
}
\right]
\end{aligned}
\end{equation}
We thus identify the two-point ``b" as
\begin{equation}\label{Eqn:b_2pt2}
\begin{aligned}
b\left({\bf x_1},{\bf x_2}\right) = -
\overline{ 
\rho'\left({\bf x_1}\right)v'\left({\bf x_2}\right)
}
\end{aligned}
\end{equation}
and thus the $\beta$ variable is
\begin{equation}\label{Eqn:Beta_b_2pt}
\begin{aligned}
\beta\left({\bf x_1},{\bf x_2}\right) = 
\frac{b\left({\bf x_1},{\bf x_2}\right)}{\overline{\rho}\left({\bf x_1}\right)} +
\frac{b\left({\bf x_2},{\bf x_1}\right)}{\overline{\rho}\left({\bf x_2}\right)}
\end{aligned}
\end{equation}

Note that the two-point $a_i$ equation may be simplified and rewritten beyond equation \ref{Eqn:a_tranport5}.  However, the current derivation is sufficient to identify the term corresponding to the BHR-3 model's $b$-term, and to illustrate the terms that couple to the mean pressure.

\section{The Two-Point ``b"-Equation}
The density equation is
\begin{eqnarray}\label{Eqn:Density}
\begin{aligned}
\frac{\partial \rho}{\partial t} + \frac{\partial \rho u_n}{\partial x_n} = 0,
\end{aligned}
\end{eqnarray}
and it's mass-weighed average is 
\begin{eqnarray}\label{Eqn:AverageDensity}
\begin{aligned}
\frac{\partial \overline{\rho}}{\partial t} + \frac{\partial \overline{\rho} \tilde{u}_n}{\partial x_n} = 0,
\end{aligned}
\end{eqnarray}
The fluctuating density is then
\begin{eqnarray}\label{Eqn:FluctDensity}
\begin{aligned}
\frac{\partial \rho'}{\partial t} + \frac{\partial \rho' \tilde{u}_n}{\partial x_n}+ \frac{\partial \rho u''_n}{\partial x_n}  = 0.
\end{aligned}
\end{eqnarray}
Using equations \ref{Eqn:AverageDensity} and \ref{Eqn:FluctDensity},
\begin{eqnarray}\label{Eqn:DensityRatio1}
\begin{aligned}
\frac{\partial }{\partial t} \left(\frac{\rho'}{\overline{\rho}}\right) &= 
\frac{1}{\overline{\rho}}\frac{\partial \rho'}{\partial t} -\frac{\rho'}{\overline{\rho}^2} \frac{\partial \overline{\rho}}{\partial t},
\end{aligned}
\end{eqnarray}
so that 
\begin{eqnarray}\label{Eqn:DensityRatio2}
\begin{aligned}
\frac{\partial \psi' }{\partial t} +\frac{\partial \psi' \tilde{u}_n}{\partial x_n} + \frac{1}{\overline{\rho}}\frac{\partial \rho u''_n}{\partial x_n}
=\psi'\frac{\partial \tilde{u}_n}{\partial x_n}
\end{aligned}
\end{eqnarray}
where
\begin{eqnarray}\label{Eqn:DensityRatio3}
\begin{aligned}
\psi &= \frac{\rho}{\overline{\rho}}, \\
\overline{\psi} &= 1, \\
\psi' &= \frac{\rho'}{\overline{\rho}}, \\
\rho &= \overline{\rho}\psi.
\end{aligned}
\end{eqnarray}
Next, note that
\begin{eqnarray}\label{Eqn:SpecVol0}
\begin{aligned}
\frac{\partial v}{\partial t} &= \frac{\partial}{\partial t}\left(\frac{1}{\rho}\right) = -\frac{1}{\rho^2}\frac{\partial \rho}{\partial t} \\
&=\frac{1}{\rho^2}\frac{\partial \rho u_n}{\partial x_n} 
= \frac{1}{\rho} \left[\frac{\partial u_n}{\partial x_n}- \rho u_n\frac{\partial}{\partial x_n}\left(\frac{1}{\rho}\right)\right]
\end{aligned}
\end{eqnarray}
so that
\begin{eqnarray}\label{Eqn:SpecVol1}
\begin{aligned}
\frac{\partial v}{\partial t} + u_n\frac{\partial v}{\partial x_n} = v\frac{\partial u_n}{\partial x_n},
\end{aligned}
\end{eqnarray}
or
\begin{eqnarray}\label{Eqn:SpecVol2}
\begin{aligned}
\frac{\partial v}{\partial t} + \frac{\partial u_n v}{\partial x_n} = 2v\frac{\partial u_n}{\partial x_n}.
\end{aligned}
\end{eqnarray}
The average specific volume is thus
\begin{eqnarray}\label{Eqn:AvgSpecVol}
\begin{aligned}
\frac{\partial \overline{v}}{\partial t} + \frac{\partial }{\partial x_n} \left(\overline{v} \tilde{u}_n +\overline{v u''_n}\right)
= 2\overline{v\frac{\partial u_n}{\partial x_n}}.
\end{aligned}
\end{eqnarray}
The fluctuating specific volume is then governed by 
\begin{eqnarray}\label{Eqn:FluctSpecVol}
\begin{aligned}
\frac{\partial v'}{\partial t} + \frac{\partial }{\partial x_n} \left(v' \tilde{u}_n +v u''_n-\overline{v u''_n} \right)
= 2\left[v\frac{\partial u_n}{\partial x_n}-\overline{v\frac{\partial u_n}{\partial x_n}}\right].
\end{aligned}
\end{eqnarray}
Equations \ref{Eqn:DensityRatio2} and \ref{Eqn:FluctSpecVol} can be used to construct an equation for 
$\overline{\psi'\left({\bf x_1}\right)v'\left({\bf x_2}\right)}$;
\begin{eqnarray}\label{Eqn:b_half1}
\begin{aligned}
\frac{\partial \overline{\psi'\left({\bf x_1}\right)v'\left({\bf x_2}\right)} }{\partial t} 
&+ \frac{\partial }{\partial x_{1n}} \left\{\overline{\psi'\left({\bf x_1}\right)v'\left({\bf x_2}\right)} \tilde{u}_n\left({\bf x_1}\right)\right\}
+ \frac{\partial }{\partial x_{2n}} \left\{\overline{\psi'\left({\bf x_1}\right)v'\left({\bf x_2}\right)} \tilde{u}_n\left({\bf x_2}\right)\right\} \\
&+\frac{\partial}{\partial x_{1n}}\left\{ \overline{v'\left({\bf x_2}\right)u''_n\left({\bf x_1}\right)}\right\}
+ 
\frac{\overline{v'\left({\bf x_2}\right)u''_n\left({\bf x_1}\right)}}{\overline{\rho}\left({\bf x_1}\right)}\frac{\partial \overline{\rho}\left({\bf x_1}\right)}{\partial x_{1n}} \\
&+\frac{\partial}{\partial x_{1n}}\left\{ \overline{\psi'\left({\bf x_1}\right)v'\left({\bf x_2}\right)u''_n\left({\bf x_1}\right)}\right\}
+ 
\frac{\overline{\psi'\left({\bf x_1}\right)v'\left({\bf x_2}\right)u''_n\left({\bf x_1}\right)}}{\overline{\rho}\left({\bf x_1}\right)}\frac{\partial \overline{\rho}\left({\bf x_1}\right)}{\partial x_{1n}} \\
&+\frac{\partial}{\partial x_{2n}}\left\{\overline{v}\left({\bf x_2}\right)\overline{\psi'\left({\bf x_1}\right) u''_n\left({\bf x_2}\right) }\right\}
+\frac{\partial}{\partial x_{2n}}\left\{\overline{\psi'\left({\bf x_1}\right) v'\left({\bf x_2}\right)u''_n\left({\bf x_2}\right) }\right\} \\
&= 2\overline{v}\left({\bf x_2}\right) \overline{ \psi'\left({\bf x_1}\right) \frac{\partial u''_n\left({\bf x_2}\right)}{\partial x_{2n}}}
+2\overline{ \psi'\left({\bf x_1}\right)v'\left({\bf x_2}\right)  \frac{\partial u''_n\left({\bf x_2}\right)}{\partial x_{2n}}} \\
&+\overline{ \psi'\left({\bf x_1}\right)v'\left({\bf x_2}\right) } \frac{\partial \tilde{u}_n\left({\bf x_1}\right)}{\partial x_{1n}}
\end{aligned}
\end{eqnarray}
The symmetrized equation is thus
\begin{equation}\label{Eqn:b_half2}
\begin{aligned}
\frac{\partial \beta\left({\bf x_1},{\bf x_2}\right) }{\partial t} 
&+ \frac{\partial }{\partial x_{1n}} \left[\beta\left({\bf x_1},{\bf x_2}\right)\tilde{u}_n\left({\bf x_1}\right)\right]
+ \frac{\partial }{\partial x_{2n}} \left[\beta\left({\bf x_1},{\bf x_2}\right) \tilde{u}_n\left({\bf x_2}\right)\right] \\ 
&+ \frac{\partial }{\partial x_{2n}} \left[\beta\left({\bf x_2},{\bf x_1}\right) \tilde{u}_n\left({\bf x_2}\right)\right]
+ \frac{\partial }{\partial x_{1n}} \left[\beta\left({\bf x_2},{\bf x_1}\right) \tilde{u}_n\left({\bf x_1}\right)\right] \\ 
&-\frac{1}{2}\left\{
\frac{\partial}{\partial x_{1n}}\left[\overline{v'\left({\bf x_2}\right)u''_n\left({\bf x_1}\right)}\right]
 +\frac{\partial}{\partial x_{2n}}\left[ \overline{v'\left({\bf x_1}\right)u''_n\left({\bf x_2}\right)}\right]
 \right\}\\ 
 &-\frac{1}{2}\left\{
 \frac{\overline{v'\left({\bf x_2}\right)u''_n\left({\bf x_1}\right)}}{\overline{\rho}\left({\bf x_1}\right)}\frac{\partial \overline{\rho}\left({\bf x_1}\right)}{\partial x_{1n}}
+ \frac{\overline{v'\left({\bf x_1}\right)u''_n\left({\bf x_2}\right)}}{\overline{\rho}\left({\bf x_2}\right)}\frac{\partial \overline{\rho}\left({\bf x_2}\right)}{\partial x_{2n}}
\right\} \\ 
&-\frac{1}{2}\left\{
\frac{\partial}{\partial x_{1n}}\left[ \overline{\psi'\left({\bf x_1}\right)v'\left({\bf x_2}\right)u''_n\left({\bf x_1}\right)}\right]
+\frac{\partial}{\partial x_{2n}}\left[ \overline{\psi'\left({\bf x_2}\right)v'\left({\bf x_1}\right)u''_n\left({\bf x_2}\right)}\right]
\right\}\\ 
&-\frac{1}{2}\left\{
\frac{\overline{\psi'\left({\bf x_1}\right)v'\left({\bf x_2}\right)u''_n\left({\bf x_1}\right)}}{\overline{\rho}\left({\bf x_1}\right)}\frac{\partial \overline{\rho}\left({\bf x_1}\right)}{\partial x_{1n}} 
+ \frac{\overline{\psi'\left({\bf x_2}\right)v'\left({\bf x_1}\right)u''_n\left({\bf x_2}\right)}}{\overline{\rho}\left({\bf x_2}\right)}\frac{\partial \overline{\rho}\left({\bf x_2}\right)}{\partial x_{2n}}
\right\} \\ 
&-\frac{1}{2}\left\{
\frac{\partial}{\partial x_{2n}}\left[\overline{v}\left({\bf x_2}\right)\overline{\psi'\left({\bf x_1}\right) u''_n\left({\bf x_2}\right) }\right]
  +\frac{\partial}{\partial x_{1n}}\left[\overline{v}\left({\bf x_1}\right)\overline{\psi'\left({\bf x_2}\right) u''_n\left({\bf x_1}\right) }\right]
  \right\} \\ 
&-\frac{1}{2}\left\{
\frac{\partial}{\partial x_{2n}}\left[\overline{\psi'\left({\bf x_1}\right) v'\left({\bf x_2}\right)u''_n\left({\bf x_2}\right) }\right]
+\frac{\partial}{\partial x_{1n}}\left[\overline{\psi'\left({\bf x_2}\right) v'\left({\bf x_1}\right)u''_n\left({\bf x_1}\right) }\right]
\right\}\\ 
&= \beta\left({\bf x_1},{\bf x_2}\right)\frac{\partial \tilde{u}_n\left({\bf x_1}\right)}{\partial x_{1n}}
+\beta\left({\bf x_2},{\bf x_1}\right) \frac{\partial \tilde{u}_n\left({\bf x_2}\right)}{\partial x_{2n}}\\ 
&-\overline{v}\left({\bf x_2}\right) \overline{ \psi'\left({\bf x_1}\right) \frac{\partial u''_n\left({\bf x_2}\right)}{\partial x_{2n}}}
-\overline{v}\left({\bf x_1}\right) \overline{ \psi'\left({\bf x_2}\right) \frac{\partial u''_n\left({\bf x_1}\right)}{\partial x_{1n}}} \\ 
&-\overline{ \psi'\left({\bf x_1}\right)v'\left({\bf x_2}\right)  \frac{\partial u''_n\left({\bf x_2}\right)}{\partial x_{2n}}}
-\overline{ \psi'\left({\bf x_2}\right)v'\left({\bf x_1}\right)  \frac{\partial u''_n\left({\bf x_1}\right)}{\partial x_{1n}}} 
\end{aligned}
\end{equation}

\section{Appendix Summary and Conclusions}
This finishes our summary of the the relevant two-point transport equations for variable density turbulence.  The equations presented may be recast in innumerable ways, but the fundamental couplings to the mean-pressure gradients are clearly represented in the above derivations.  A next-step in developing a model will be to recast the above equations using the relative, ${\bf r} =\left({\bf x_1} - {\bf x_2}\right)$  and centered ${\bf x_c} =\left({\bf x_1} + {\bf x_2}\right)/2$ coordinates used by Bernard {\it et al} \cite{BHRZ1}.  Although Clark and Spitz \cite{CSModel} and Steinkamp {\it et al.} \cite{Steinkamp} choose to pursue a Fourier transformation of the the resultant two-point equations, it seems well-worn exploring the potential for modeling the transport equations in configurational space, thus avoiding the issues regarding the use of Fourier representations in inhomogeneous and compressible circumstances.  However, such an endeavor is currently beyond the scope of the current project.


\end{document}